\documentclass[a4paper,11pt]{article}
\usepackage{jheppub}
\usepackage[T1]{fontenc}
\usepackage{graphicx}
\usepackage{amsmath}
\usepackage{amssymb}
\usepackage{amsfonts}
\usepackage{bm}
\usepackage[table]{xcolor}

\def\be{\begin{equation}}
\def\ee{\end{equation}}
\def\bea{\begin{eqnarray}}
\def\eea{\end{eqnarray}}

\newcommand{\mc}[1]{\mathcal{#1}}
\newcommand{\f}[2]{\frac{#1}{#2}}
\title{\boldmath Cosmology in theories with derivative matter coupling}

\author[~]{Zahra Haghani and}
\author{Shahab Shahidi}

\affiliation[~]{School of Physics, Damghan University, Damghan, 
	41167-36716, Iran.}
\emailAdd{z.haghani@du.ac.ir}
\emailAdd{s.shahidi@du.ac.ir}

\abstract{
	A new class of modified gravity theories with a healthy higher order derivative terms of a function of the matter Lagrangian $f(L_m)$ is considered. Generally the energy momentum tensor is not conserved, leading to the fifth force similar to $f(R,T)$ theories. We will however show that in the FRW background there exists two possibilities corresponding to conservative and non-conservative cases. Cosmological implications of both cases with different functions of the matter Lagrangian $f$ will be investigated in details and we will show that current observational data can be satisfied in this model. The non-conservative case however, predicts less matter sources at early times and more deceleration. Evolution of the matter density perturbation in the longitudinal gauge is also considered for dust matter sources. We will show that the late time behavior of the matter density perturbation is in agreement with current observational data at sub-horizon limit.
}
\begin{document}
	\maketitle

\section{Introduction}
The discovery of the late time accelerated expansion of the Universe \cite{expand} which can not be explained in the standard theory of general relativity, has motivated scientists to search for modified theories of gravity. Perhaps the simplest modification is to add to the Einstein's theory, a positive cosmological constant. However, this arise the cosmological constant problem \cite{ccprob} encouraging cosmologists to search for a modified theory with an additional dynamical field.

Modification of the Einstein gravity action can be done in many ways, including the addition of some new dynamical fields to the theory or modifying the gravitational field itself. In the first class, new dynamical fields can be considered as a part of geometry, or as a part of the matter sector. As a result, one can add an additional scalar/vector field to the theory in such a way that the resulting theory does not contain more than second order derivatives in the fields. This ensures that the modified theory is free from Ostrogradski/gradient instabilities \cite{inst}. Also, the tensor mode of the theory should propagate with the speed of light, demanded by recent observations of the gravitational waves \cite{gw,prls}. Imposing all of these constraints, a few number of theories survive, including Brans-Dicke theory \cite{bd}, Quintessence/k-essence \cite{quint}, some special classes of the GLPV theory \cite{glpv}, a special case of DHOST \cite{dhost}, etc.. Cosmological implications of these theories are widely analyzed in the literature \cite{cosgali}.
The second class of modified gravity theories can be obtained by enriching the gravitational field itself. This include $f(R)$ theories \cite{fR}, Weyl, Cartan \cite{weyl}, Weitzenbock \cite{weit} theories as well as massive gravity theories \cite{massive}, etc. \cite{other}. It should be mentioned that most of theories in the second class can be written as a gravity theory with some additional dynamical fields, by some suitable field redefinition. One of the famous examples is the $f(R)$ theory of gravity which is equivalent to the Brans-Dicke theory with vanishing kinetic term of the Brans-Dicke scalar field \cite{bd}. This shows that the two classes mentioned above are not logically distinct.
However, the common thing about all above modifications is that they add some degrees of freedom to the theory.

Another way to think about modification of the gravity action is to note that in the original Einstein's theory, the matter and geometry appears in a same footing in the level of action (and also in the level of field equations). It will become more democratic if one think about other modifications of the Einstein's theory resulting from the matter sector itself. One can then generalize the Einstein's general relativity and write an action which includes an arbitrary function of the Ricci scalar and the trace of the energy momentum tensor, known as the $f(R,T)$ theories \cite{fRT}. In this theory, the matter energy momentum tensor is no longer conserved and as a result we have a fifth force which will affect the dynamics of the Universe. This represents a non-minimal coupling between matter and geometry. However matter does not represent only with the trace of the energy momentum tensor. Other classes of theories can be formed by using the matter Lagrangian $L_m$ instead of the trace of the energy momentum tensor \cite{Lmtheory}. One can also go further and assume a more complicated matter-geometry couplings like $R_{\mu\nu}T^{\mu\nu}$ \cite{ourRT}. Higher order matter couplings with Lagrangian density $\sqrt{-g}(R+f(T_{\mu\nu}T^{\mu\nu}))$ \cite{roshan,powered} can also be considered. In all of the above modified theories the matter energy momentum tensor in general is not conserved and the extra force would appear. Using the solar system tests, the extra force will restrict the form of the matter coupling in the above theories \cite{will}. 
The main consequence of the idea of adding some non-minimal couplings between matter and geometry is that it is sufficient to consider only the ordinary baryonic matter field as a matter source of the Universe. The accelerated expansion of the Universe, together with other observational data can be explained from the additional non-minimal couplings. As a result, there is no need to add other dynamical degrees of freedom to the theory.

Another way to produce non-minimal coupling between matter and geometry, is to add some derivative matter terms to the theory. This will add some couplings between matter sources and the Levi-Civita connection, instead of the Riemann tensor, through covariant derivatives of the energy momentum tensor.
Derivative matter coupling is considered in the level of the equation of motion in \cite{harko}. The authors wrote the most general combinations of the matter energy momentum tensor up to second order in derivatives in such a way that the result becomes divergence free. This implies that point particles move along the geodesic lines and no extra fifth force appears in the theory. Cosmological implications of the mode was also investigated in details. The problem with the approach presented in \cite{harko} is that the equation of motion can not be obtained from variational principle of an action.

A realistic idea would be to consider derivative matter couplings in the level of action. This is the main idea of the present paper. The coupling can be constructed by matter Lagrangian $\mc{L}_m$ or the trace of the energy momentum tensor $T^\mu_\mu$. In this paper, we will use the matter Lagrangian to construct new couplings, and try to build a healthy derivative self-interactions of the matter Lagrangian. By considering the fact that the matter Lagrangian is a scalar field, it will be sufficient to choose an appropriate higher derivative Lagrangian of a scalar field to have a healthy derivative matter couplings. As was discussed above the beyond Horndeski family would be a good choice since it is free from the ghost instabilities and also satisfies the gravitational wave observations \cite{gw}. So, we will consider a subclass of the Horndeski theory where the scalar field is now the matter Lagrangian. Specifically in this paper, we will consider the second and third Galileon interaction terms as a derivative interaction of the matter Lagrangian. These two terms has a property that they are widely investigated in the literature, and satisfies the recent constraint on the speed of the gravitational wave. The resulting field equation will contain derivative terms of the matter Lagrangian which will be responsible for the late time accelerated expansion of the Universe. We will see that the matter energy momentum tensor is not conserved in general and the fifth force will be appeared in the theory. The fifth force however has a property that over the FRW ansatz, there is a possibility that the matter field is conserved. So, the theory has a reach cosmological consequences which we will investigate in details.

The plan of the paper is as follows: In the next section, we will introduce the action and obtain the field equations together with the conservation equation of the matter field. In section \ref{cos}, we will adopt various functions of the matter Lagrangian and investigate the cosmological consequences of the theory in both conservative and non-conservative cases. In section \ref{matden}, we will obtain the evolution equations of the matter density perturbations and discuss the deep sub-horizon behavior of the model. Section \ref{con} will be devoted to final discussions and conclusions.
\section{The model}
In this paper, we are going to consider the effects of the derivative couplings of the matter Lagrangian $L_m$ in the dynamics of the Universe. Note that the matter Lagrangian is a scalar field and it will be better to use a healthy scalar theory to make a matter coupling. One of the most interesting healthy higher derivative scalar field theories is the Horndeski theory. As was discussed in the introduction, the fourth and fifth Horndeski Lagrangians are constrained by the recent gravitational wave observations. In this paper, for the sake of simplicity, we will only consider the second and the third Galileon interactions which obviously satisfied the above observations \cite{prls}. Let us then consider an action functional of the form
\begin{align}\label{action}
S=\int d^4x\sqrt{-g}\bigg[\kappa^2 R+\alpha\,\nabla_\mu f\,\nabla^\mu f+\beta \,\Box f \,\nabla_\mu f\,\nabla^\mu f\bigg]+S_m,
\end{align}
where $f=f(L_m)$ is an arbitrary function of the matter Lagrangian and $S_m$ is the ordinary matter action. The last two terms are the Galileon Lagrangians for the scalar field $f$ and $\alpha$ and $\beta$ are twp arbitrary constants.

Using the definition of the energy momentum tensor
\begin{align}
T_{\mu\nu}=\f{-2}{\sqrt{-g}}\f{\delta(\sqrt{-g}L_m)}{\delta g^{\mu\nu}},
\end{align}
and the variation of the matter Lagrangian
\begin{align}
\delta L_m=\f{1}{2}\big(L_m g_{\mu\nu}-T_{\mu\nu}\big)\delta g^{\mu\nu},
\end{align}
one can obtain the equation of motion of the metric field as
\begin{align}\label{geom}
\kappa^2G_{\mu\nu}&-\f12T_{\mu\nu}+f^\prime\left(T_{\mu\nu}-g_{\mu\nu}L_m \right)\left(\alpha \,\Box f+\beta\,Q\right)+\alpha\left(\nabla_\mu f\nabla_\nu f-\f12g_{\mu\nu}\nabla_\alpha f\nabla^\alpha f \right)\nonumber\\&+\beta\bigg(g_{\mu\nu}\,f\nabla_\alpha\nabla_\beta f\nabla^\alpha\nabla^\beta f+\Box f\, \nabla_\mu f \nabla_\nu f-\nabla^\alpha f \,\nabla_\alpha\left(\nabla_{\mu}f\nabla_{\nu}f\right) \bigg)=0,
\end{align}
where prime denotes derivative with respect to the argument and we have defined
\begin{align}
Q=\left(\Box f\right)^2-R_{\alpha\beta}\nabla^{\alpha}f\nabla^{\beta}f-\nabla_{\alpha}\nabla_{\beta}f \nabla^{\alpha}\nabla^{\beta}f.
\end{align}
The conservation of the energy momentum tensor can be obtained as
\begin{align}\label{conser1}
\Big(1-
f^\prime\left(\alpha\,\Box f +\beta \,Q\right)\Big)\nabla^\mu &T_{\mu\nu}=2(L_mg_{\mu\nu}-T_{\mu\nu})\Big[\beta \,\nabla^\mu \left(Q f^\prime\right)+\alpha\, \nabla^\mu\left(f^\prime \Box f\right)\Big].
\end{align}
The above conservation equation could be simplified in the cases where the matter field can be described by a perfect fluid. Suppose that the Lagrangian density and the energy momentum tensor can be written as
$$L_m=-\rho,\quad T_{\mu\nu}=(\rho+p)u_\mu u_\nu+pg_{\mu\nu},$$
where $\rho$ and $p$ are the energy density and thermodynamics pressure of the fluid, and $u_\mu$ is the 4-velocity of the fluid.
In this case equation \eqref{conser1} will be simplified to
\begin{align}\label{conser2}
\Big(1-
f^\prime\left(\alpha\,\Box f +\beta \,Q\right)\Big)\nabla^\mu T_{\mu\nu}=2(\rho+p)h_{\mu\nu}\,\nabla^\mu\Big[\alpha\, f^\prime \Box f+\beta Q f^\prime\Big].
\end{align}
The factor $h_{\mu\nu}=u_\mu u_\nu+g_{\mu\nu}$ is the projection tensor into the hyper-surfaces orthogonal to $u_\mu$. In the case of FRW space time, which we will consider in the following, RHS of equation \eqref{conser1} vanishes since all the quantities depends only on time and the covariant derivative becomes orthogonal to $h_{\mu\nu}$. As a result we have two possibilities corresponding to the conservative and non-conservative matter content. In the following, we will consider both cases.

\section{Cosmological implications}\label{cos}
In this section, we will consider cosmological implications of the model \eqref{action}. Suppose that the Universe is described by the flat FRW ansatz
\begin{align}
ds^2=-dt^2+a^2(dx^2+dy^2+dz^2),
\end{align}
filled with a perfect fluid with matter Lagrangian of the form $L_m=-\rho$ and the energy momentum tensor
\begin{align}\label{pert}
T^\mu_\nu=\textmd{diag}(-\rho,p,p,p),
\end{align} 
where $a(t)$ is the scalar factor. Also, $\rho$ and $p$ are functions of time. Substituting in \eqref{geom}, one can obtain the Friedman and Raychaudhuri equations as
\begin{align}\label{frid}
&6\kappa^2H^2=\rho-\alpha f^{\prime2}\,\dot{\rho}^2-6\beta‌ H\,f^{\prime3}\,\dot{\rho}^3,
\end{align}
and
\begin{align}\label{ray}
-4\kappa^2\dot{H}&=\rho+p+2f^{\prime}(\rho+p)(\alpha+6\beta f^{\prime} H\dot{\rho})(f^{\prime\prime}\dot{\rho}^2-f^{\prime}\ddot{\rho})-2\beta f^{\prime3}\dot{\rho}^2(3H\dot{\rho}-\ddot{\rho})\nonumber\\
&-6\beta (\rho+p)(\dot{H}+3H^2)f^{\prime3}\dot{\rho}^2-6\alpha H(\rho+p)f^{\prime2}\dot{\rho}-2(\alpha+\beta f^{\prime\prime}\dot{\rho}^2)f^{\prime2}\dot{\rho}^2,
\end{align}
where dot denotes time derivative, $H=\dot{a}/a$ is the Hubble parameter and $f^{\prime}=\partial f/\partial L_m$ and $f^{\prime\prime}=\partial^2 f/\partial L_m^2$. In order to close the above system, we will also assume that the perfect fluid obeys the barotropic equation of state, $p=\omega\rho$. The energy momentum conservation equation can be written as
\begin{align}\label{cons2}
\Big(3 H (p+\rho )+\dot{\rho }\Big)& \Bigg(2 f^{\prime2} \bigg(3 H \dot{\rho }
\left(\alpha-2 \beta f^{\prime\prime}\dot{\rho }^2\right)+\alpha \ddot{\rho
}\bigg)\nonumber\\&-2 \alpha f^\prime f^{\prime\prime} \dot{\rho }^2+6 \beta f^{\prime3}
\dot{\rho } \left((3 H^2+\dot{H}) \dot{\rho }+2 H \ddot{\rho
}\right)-1\Bigg)=0.
\end{align}
As one can see from the above conservation equation, there are two possibilities for this model. The first one which obtains by vanishing of first parenthesis in \eqref{cons2} corresponds to the conservative cosmology. Also if the second parenthesis in \eqref{cons2} vanishes, the matter is not conserved. We will investigate both cases in the following.
\subsection{Conservative cosmology}
In this section, we will consider cosmological implications of the conservative case. Suppose that the baryonic matter content of the Universe consists of both radiation with $p_r=\rho_r/3$ and dust with $p_m=0$. We introduce a redshift $z$ as an independent variable defined as $1+z=1/a$. Therefore for a general function of time $g$ we have
$$\f{dg}{dt}=-(1+z)H\f{dg}{dz}.$$
Moreover, we defined dimensionless quantities as
\begin{align}\label{dim1}
H=H_0h(z),\quad \Omega_i=\f{\rho_i}{\rho_c},
\end{align}
where $i=rad,dust$,  $\rho_c=6\kappa^2H_0^2$ is the current critical density of the Universe and $H_0$ is the current Hubble parameter. Also as an indicator of accelerated expansion of the Universe, we will use the deceleration parameter defined as
\begin{align}
q=-1+\f{d}{dt}\f{1}{H(t)}=-1+(1+z)\f{1}{H(z)}\f{d H(z)}{dz}.
\end{align}
From the above definitions, one can solve the conservation of the energy momentum tensor $\dot{\rho}+3H(\rho+p)=0$ for matter and radiation to obtain
\begin{align}\label{cons}
\Omega_r=\Omega_{r0}(1+z)^4,\qquad \Omega_m=\Omega_{m0}(1+z)^3,
\end{align}
where $\Omega_{r0}=1.68\times10^{-5}$ and $\Omega_{m0}=0.301$ are the current density parameters \cite{current}.
Let us analyze some special choices for the function $f$.
\subsubsection{The power-law case: $f=\gamma (-L_m/\rho_c)^n$}
\begin{figure}[h]\label{fig1}
	\centering
	\includegraphics[scale=0.446]{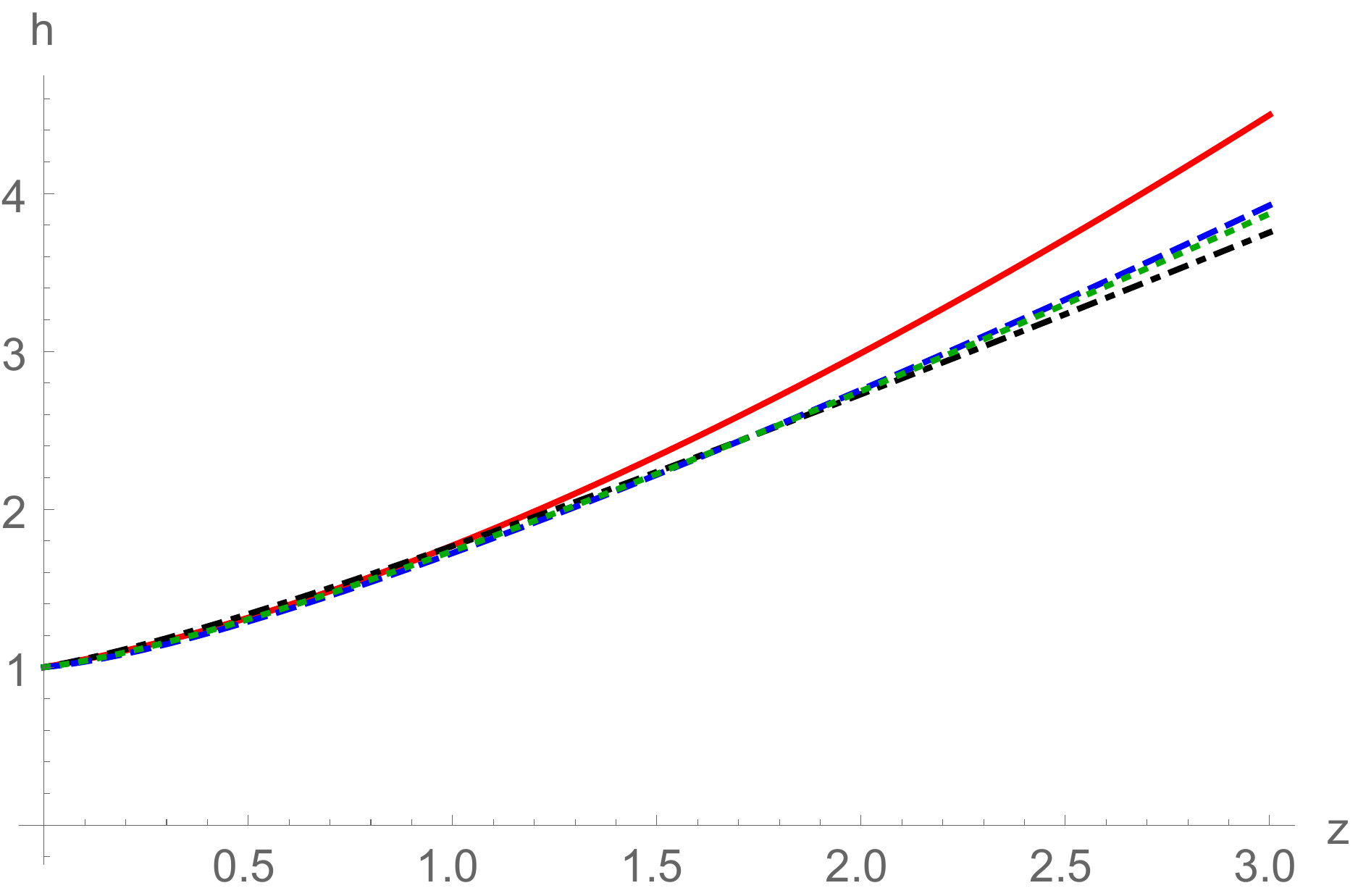}
	\includegraphics[scale=0.446]{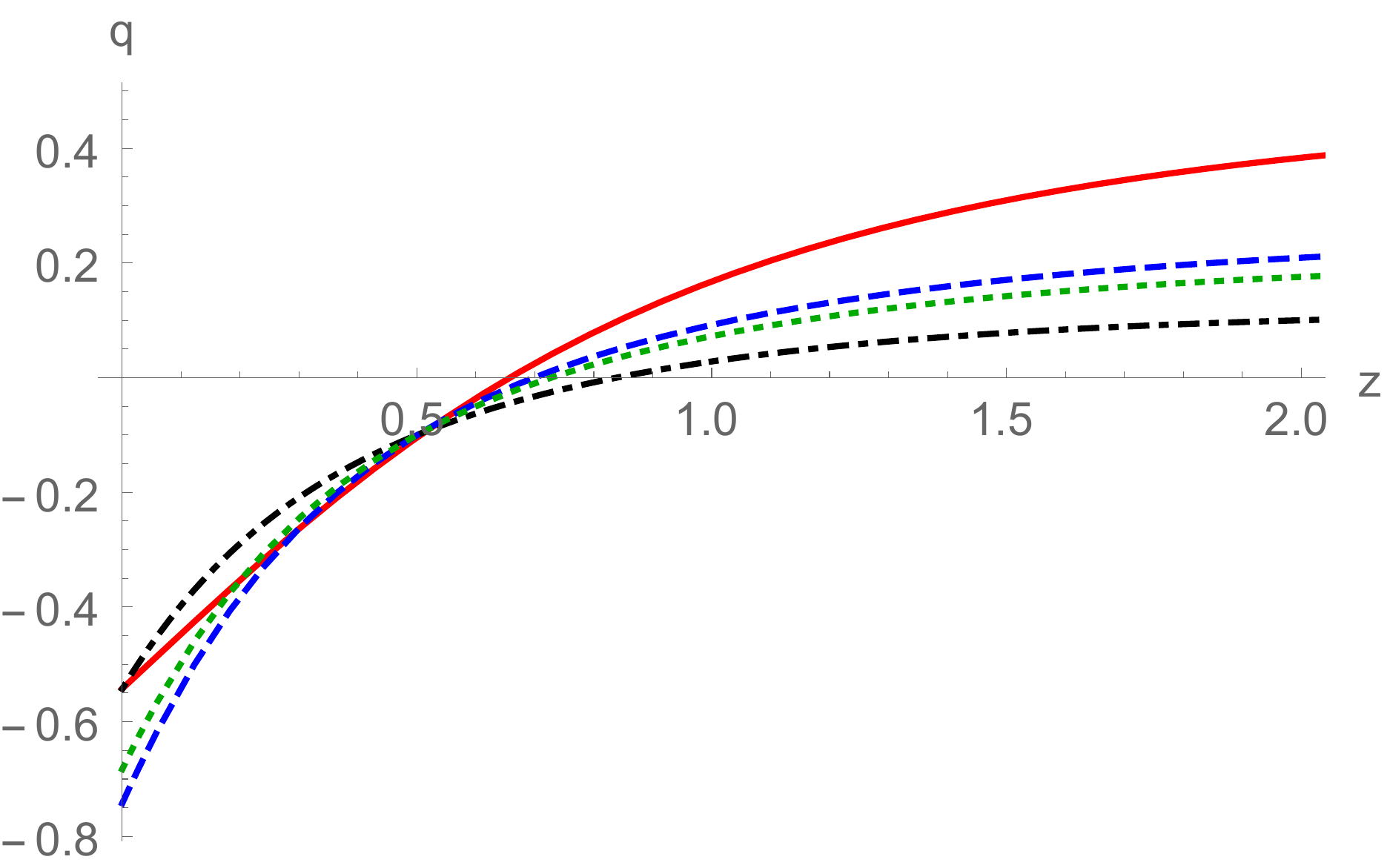}
	\caption{The variation of Hubble and deceleration parameters as functions of the redshift $z$ for $n=-0.12$ (dot-dashed), $n=-0.16$ (dotted) and $n=-0.18$ (dashed) with $\epsilon=-3.12,-1.34,-0.96$ and $\eta=-5.34,-2.84,-2.19$ respectively, for power-law function $f\propto (L_m)^n$.}
\end{figure}

Let us consider the power-law case $f=\gamma (-L_m/\rho_c)^n$ where $\gamma$ and $n$ are arbitrary constants. In this case the Friedman equation takes the form
\begin{align}
h^2=\Omega-n^2\Omega^{2(n-1)}\dot{\Omega}^2(\eta-n\epsilon h\Omega^{n-1}\dot{\Omega}),
\end{align}
where $\Omega\equiv\Omega_m+\Omega_r$ and we have defined dimensionless constants
\begin{align}\label{constants}
\eta=\f{\gamma^2\alpha}{6\kappa^2},\qquad\epsilon=\f{\beta\gamma^3 H_0^2}{\kappa^2}.
\end{align}
Transforming to the redshift coordinates and using the conservation equation \eqref{cons}, one obtains
\begin{align}
\epsilon n^3(1&+z)^{9n}(3\omega_0+(1+z)\Omega_{r0})^3\omega_0^{3n-1}h(z)^4\nonumber\\&+\big(1+\eta n^2(1+z)^{6n}(3\omega_0+(1+z)\Omega_{r0})^2\omega_0^{2n-1}\big)h(z)^2-\omega_{0}(1+z)^3=0,
\end{align}
where we have introduced $\omega_0(z)=\Omega_{m0}+(1+z)\Omega_{r0}$. In figure \eqref{fig1} we have depicted the variation of Hubble and  deceleration parameters as functions of $z$ for $n=-0.12$ (dot-dashed), $n=-0.16$ (dotted) and $n=-0.18$ (dashed) with $\epsilon=-3.12,-1.34,-0.96$ and $\eta=-5.34,-2.84,-2.19$ respectively. The red solid line is the $\Lambda$CDM curve. It is worth mentioning that in the conservative case, only the values $-1<n<0$ can be fitted with observational data. The other values of $n$ will make more acceleration at late times, incompatible with observations. Also one can see that the Universe has a smaller deceleration at early times compared to the standard $\Lambda$CDM model. This can be explained by noting that with $n<0$, the energy density in derivative terms has a negative power and at early times produces more acceleration.
\subsubsection{The case: $f=\gamma(-L_m/\rho_c)^n\exp(-\zeta L_m/\rho_c)$}
\begin{figure}[h]\label{fig2}
	\centering
	\includegraphics[scale=0.446]{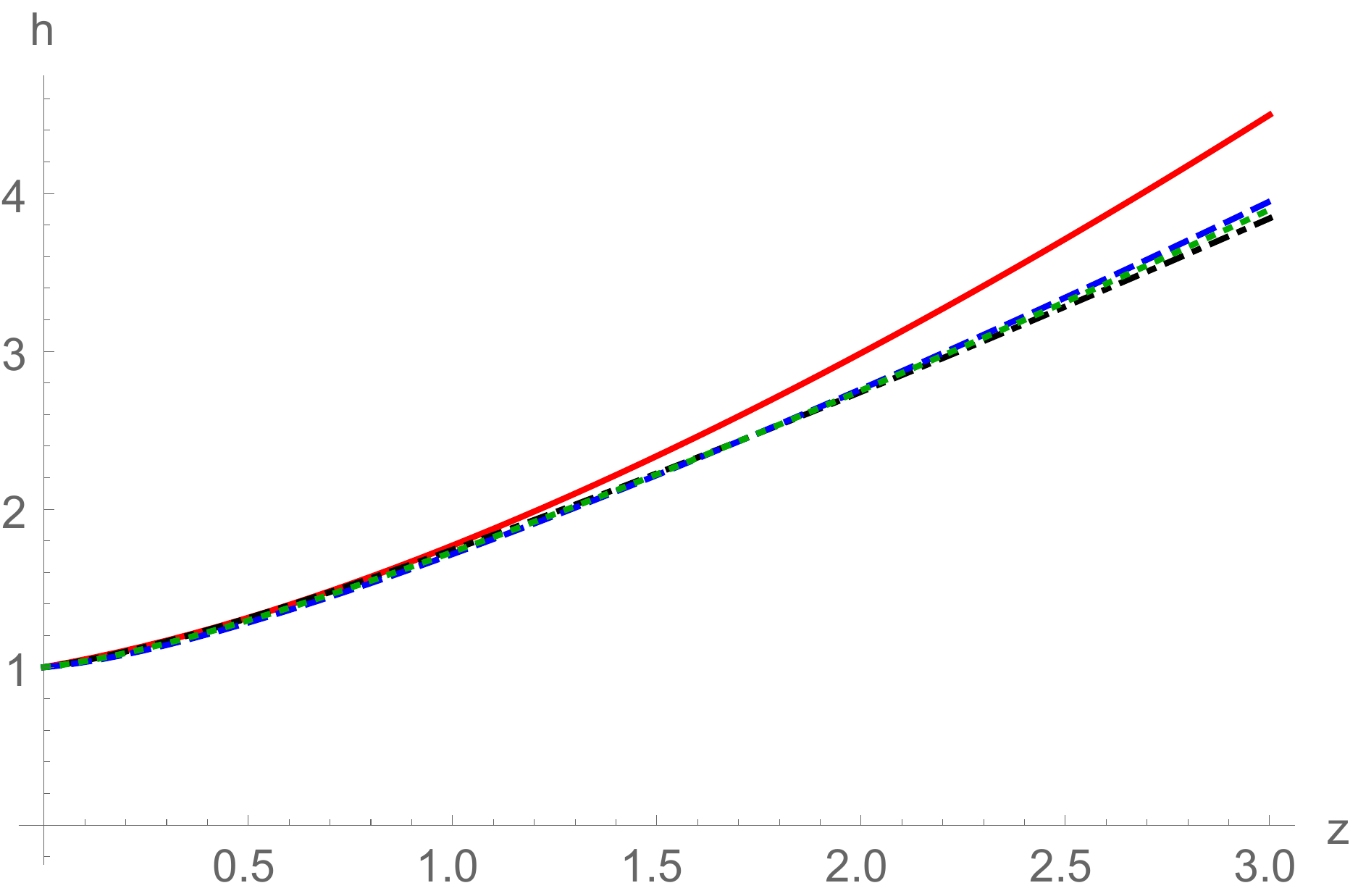}
	\includegraphics[scale=0.446]{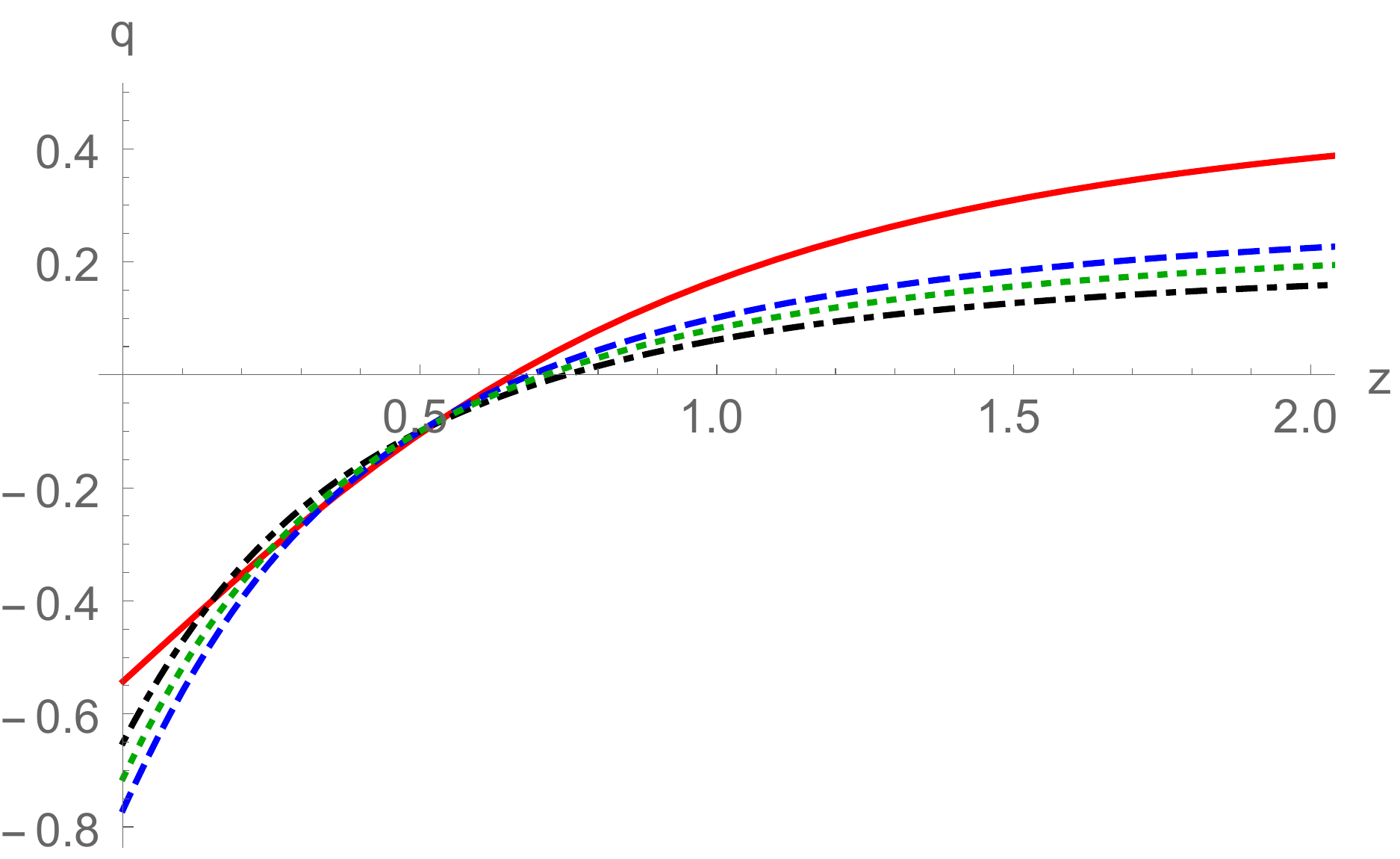}
	\caption{The variation of Hubble and deceleration parameters as functions of the redshift $z$ for $n=-0.15$ (dot-dashed), $n=-0.17$ (dotted) and $n=-0.19$ (dashed) with $\zeta=10^{-5}$ and $\epsilon=-1.62,-1.13,-0.82$ and $\eta=-3.28,-2.49,-1.95$ respectively, for the function $f\propto L_m^n\exp(L_m)$.}
\end{figure}

In this case the Friedman equation takes the form
\begin{align}
h^2=\Omega-\eta e^{2\zeta\Omega}\Omega^{2(n-1)}\dot{\Omega}^2(n+\zeta\Omega)^2+\epsilon h e^{3\zeta\Omega}\Omega^{3(n-1)}\dot{\Omega}^3(n+\zeta\Omega)^3,
\end{align}
where $\eta$ and $\epsilon$ are defined in \eqref{constants} and $\zeta$ is a constant. Using the conservation equations \eqref{cons} and transforming to the redshift coordinates, one can see that the Hubble parameter should be obtained from the following algebraic equation
\begin{align}
\Omega_0 ^4-\Omega_0 \left(\eta \Omega_1^2 e^{2 \zeta  \Omega_0 } \Omega_0 ^{2 n}
(\zeta \Omega_0 +n)^2+\Omega_0^2\right) h(z)^2 - \epsilon \Omega_1^3 e^{3 \zeta
	\Omega_0 }\Omega_0 ^{3 n} (\zeta  \Omega_0 +n)^3h(z)^4=0.
\end{align}
Here we have introduced 
\begin{align}\label{new}
\Omega_0(z)=(1+z)^3(\Omega_{m0}+(1+z)\Omega_{r0}),\qquad \Omega_1(z)=(1+z)^3(3\Omega_{m0}+4\Omega_{r0}(1+z)).
\end{align}
In figure \eqref{fig2}, we have depicted the Hubble and deceleration parameters as functions of the redshift. 
It should be noted that the exponential term will cause very rapid acceleration at late times. As a result, we assume a small value for $\zeta=10^{-5}$. The qualitative behavior of the figures is the same as in the power-law case, \eqref{fig1}. However, one can see that the exponential term makes the Universe to accelerate faster. It should also be emphasized that the best fit with the observational data could be found for values of $n$ in the range $(-1,0)$. Other values of $n$ will produce more rapid acceleration compared to the $\Lambda$CDM data (red curves in the figures).
\subsubsection{The case: $f=\gamma(-L_m/\rho_c)^n\exp(-\zeta \rho_c/L_m)$}
\begin{figure}[h]\label{fig3}
	\centering
	\includegraphics[scale=0.446]{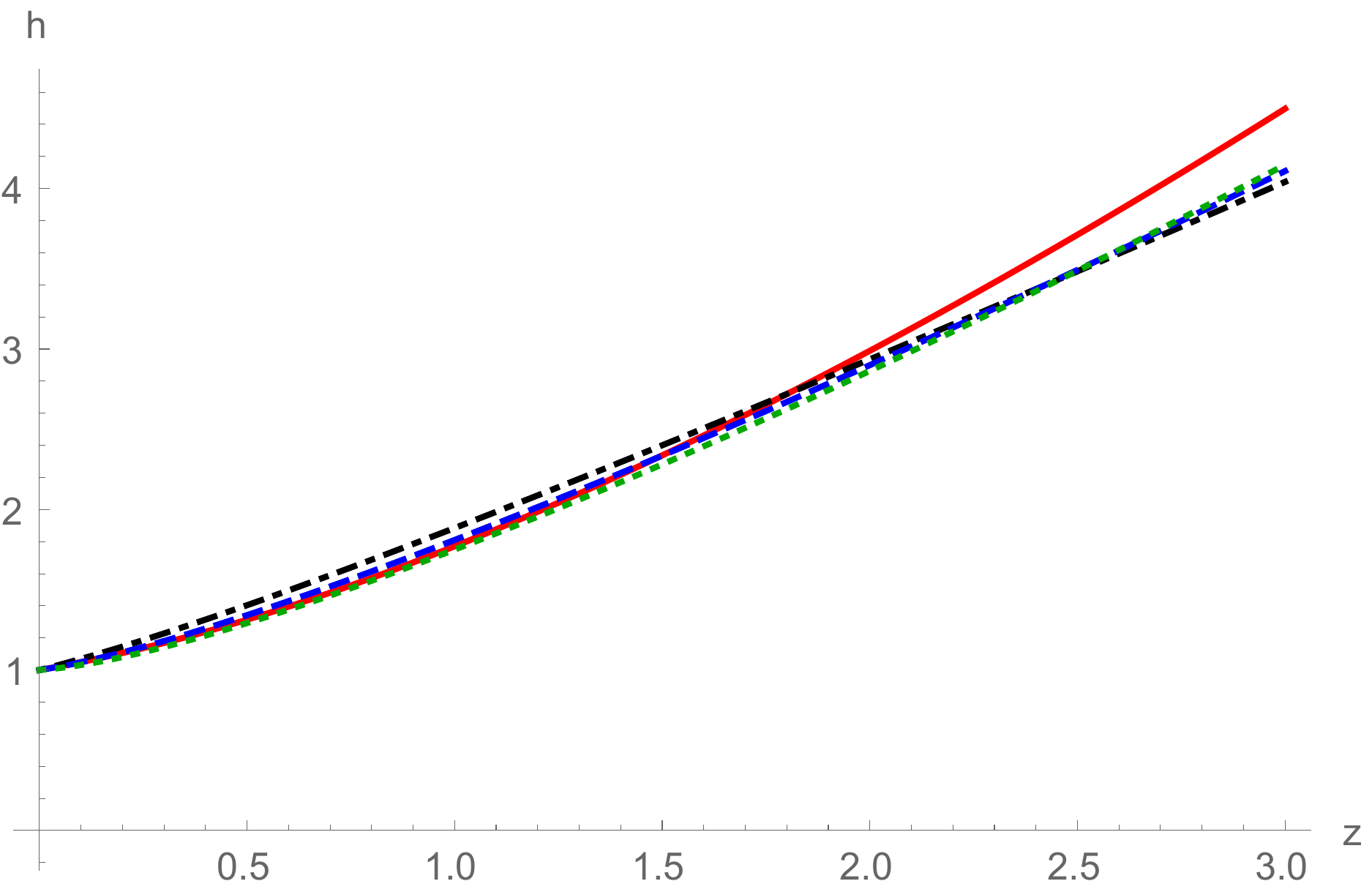}
	\includegraphics[scale=0.446]{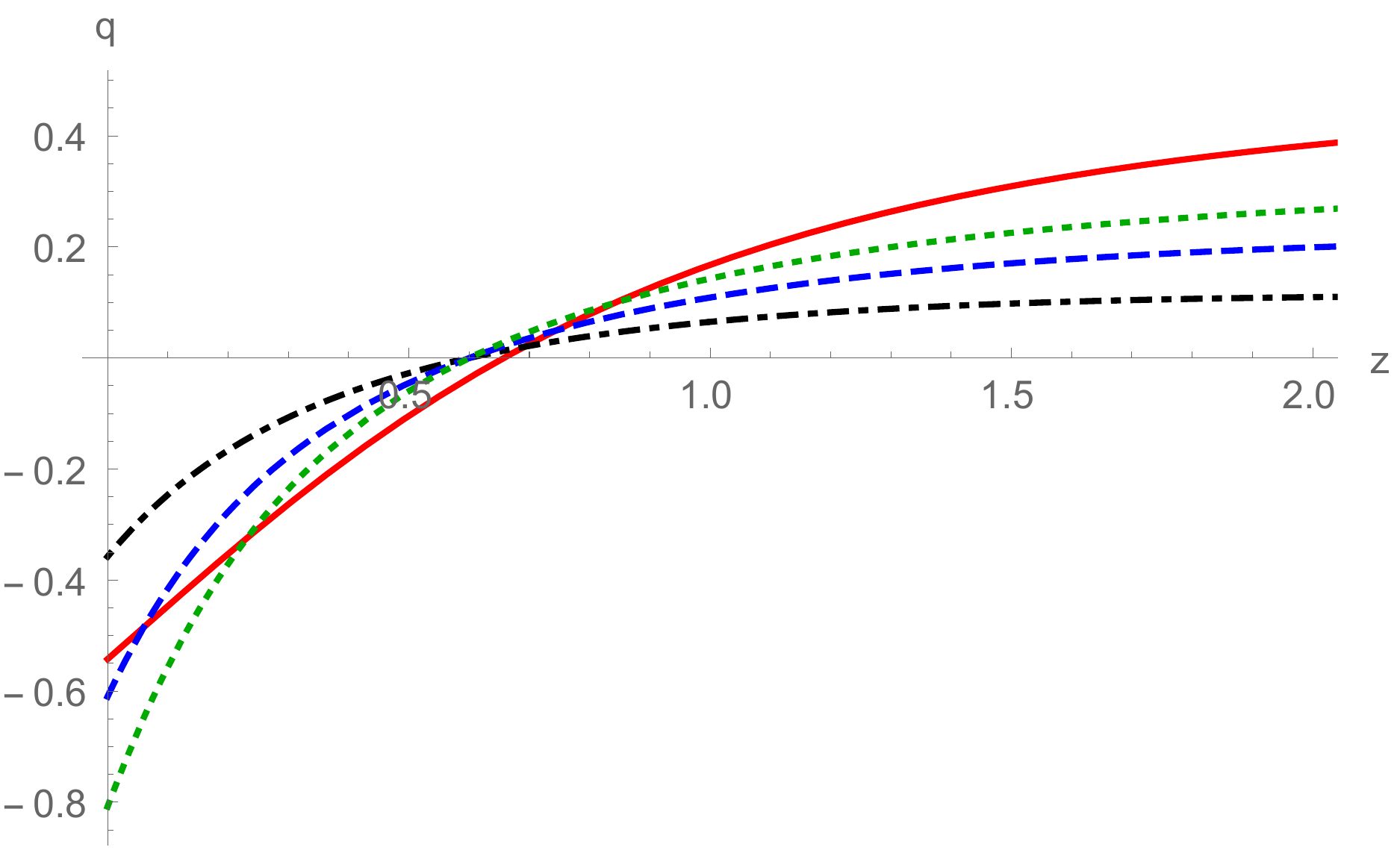}
	\caption{The variation of Hubble and deceleration parameters as functions of the redshift $z$ for $n=-0.10$ (dot-dashed), $n=-0.15$ (dashed) and $n=-0.20$ (dotted) with $\zeta=10^{-4}$ and $\epsilon=-2.94,-0.94,-0.46$ and $\eta=-7.14,2.93,-1.55$ respectively, for the function $f\propto L_m^n\exp(1/L_m)$.}
\end{figure}

The Friedman equation in this case can be written as
\begin{align}
h^2=\Omega-\eta e^{\f{2\zeta}{\Omega}}\Omega^{2(n-2)}(\zeta-n\Omega)^2\dot{\Omega}^2-\epsilon h e^{\f{3\zeta}{\Omega}}\Omega^{3(n-2)}(\zeta-n\Omega)^3\dot{\Omega}^3,
\end{align}
where $\epsilon$, $\eta$ are defined in \eqref{constants} and $\zeta$ is a constant.  Using conservation equations \eqref{cons} and transforming to the redshift coordinates, one obtains
\begin{align}
\Omega_0 ^7-\Omega_0 ^2  \left(\eta  \Omega_1^2 e^{2 \zeta /\Omega_0} \Omega_0
^{2 n} (\zeta -n\Omega_0)^2+\Omega_0 ^4\right)h(z)^2+\epsilon \Omega_1^3  e^{3
		\zeta/\Omega_0} \Omega_0^{3 n} (\zeta -n \Omega_0)^3 h(z)^4=0.
\end{align}
The definition of $\Omega_0$ and $\Omega_1$ has been made \eqref{new}. In figure \eqref{fig3}, we have plotted the Hubble and deceleration parameters as functions of the redshift.
One can see from these figures that the early time behavior is similar to the exponential case of the previous subsection. However, at late times where the energy density decreases, the behavior of the inverse-exponential function differs from the exponential one. One can see that for smaller value of $n$ one obtains less acceleration. As before, the best fit with observational data can be made only for the range $n\in(-1,0)$.
\subsection{Non-conservative cosmology}
\begin{figure}[h]\label{fig4}
	\centering
	\includegraphics[scale=0.446]{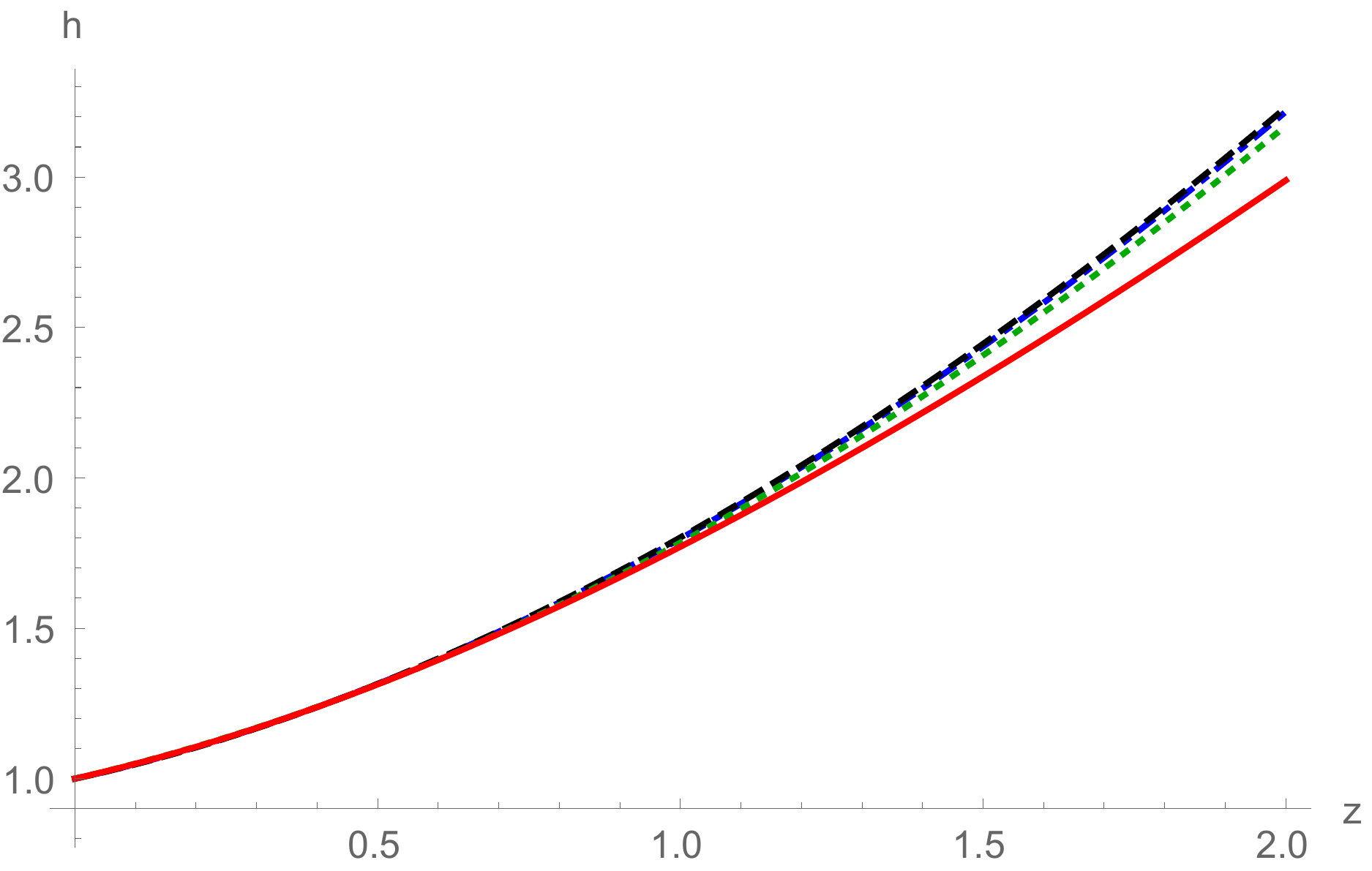}
	\includegraphics[scale=0.446]{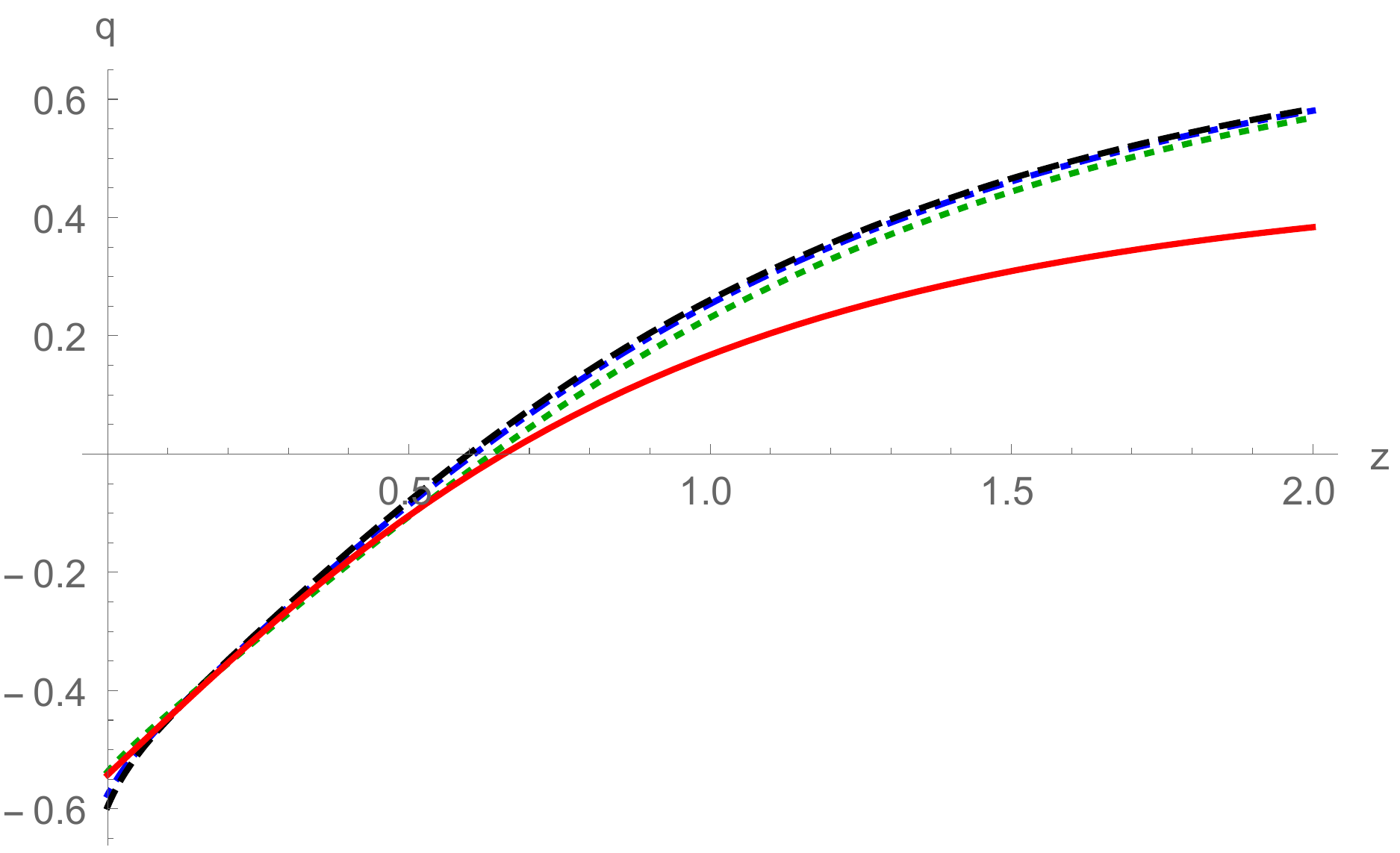}
	\caption{The variation of Hubble and deceleration parameters as functions of the redshift $z$ for $n=2.5$ (dotted), $n=3.5$ (dot-dashed) and $n=4.5$ (dashed) with $\eta=2.6,14.0,32.5$ and $\epsilon=-22.6,-269.4,-941.7$ respectively, for the power-law function $f\propto (L_m)^n$.}
\end{figure}

\begin{figure}[h]\label{fig5}
 	\centering
 	\includegraphics[scale=0.45]{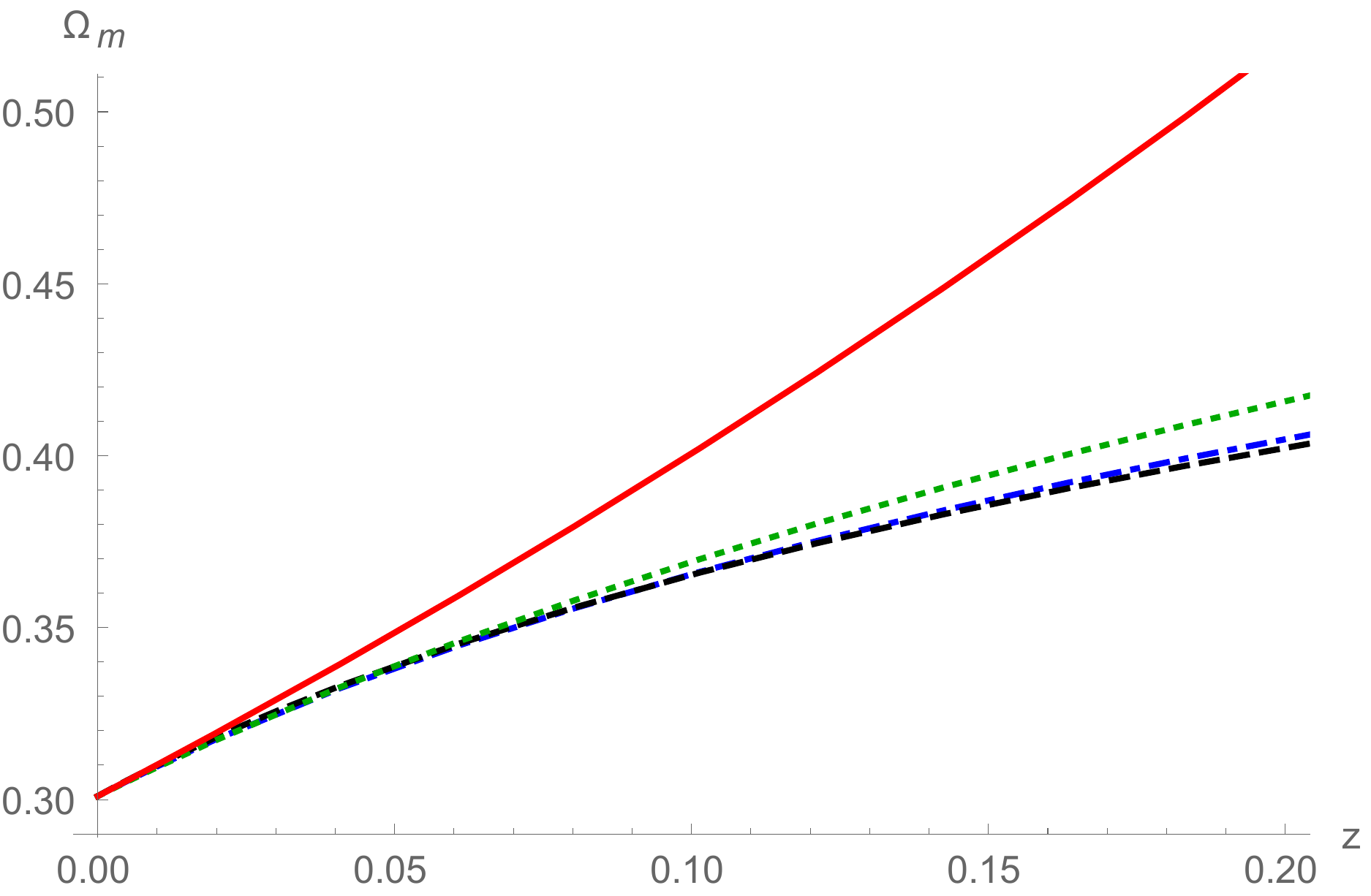}
 	\caption{The variation of the density parameter as functions of the redshift $z$ for $n=2.5$ (dotted), $n=3.5$ (dot-dashed) and $n=4.5$ (dashed) with $\eta=2.6,14.0,32.5$ and $\epsilon=-22.6,-269.4,-941.7$ respectively, for the power-law function $f\propto (L_m)^n$.}
 \end{figure}

let us now consider the case where the energy momentum tensor is not conserved. In this case, from equation \eqref{cons2}, one has
\begin{align}
2 f^{\prime2} \bigg(3 H \dot{\rho }
\left(\alpha-2 \beta f^{\prime\prime}\dot{\rho }^2\right)+\alpha \ddot{\rho
}\bigg)-2 \alpha f^\prime f^{\prime\prime} \dot{\rho }^2+6 \beta f^{\prime3}
\dot{\rho } \left((3 H^2+\dot{H}) \dot{\rho }+2 H \ddot{\rho
}\right)=1
\end{align}
Let us now consider a dust dominated Universe with power-law function $f=\gamma(-L_m/\rho_c)^n$ where $\gamma$ and $n$ are some constants. One obtains the dimensionless Friedman equation as
\begin{align}
h^2=\Omega_m-n^2\Omega_m^{2(n-1)}\dot{\Omega}_m^2(\eta-n\epsilon h\Omega_m^{n-1}\dot{\Omega}_m).
\end{align}
Also, the (non)-conservation equation of the energy momentum tensor \eqref{cons2} can be written as
\begin{align}
\Omega_m^4&-2 \eta  n^2 \Omega_m^{2 n+1} \left(3 h \Omega_m\dot{\Omega}_m+(n-1) \dot{\Omega}_m^2+\Omega_m
\ddot{\Omega}_m\right)\nonumber\\&+n^3 \epsilon  \Omega_m^{3 n} \dot{\Omega}_m \left(\dot{\Omega}_m\left(\Omega_m \left(\dot{h}+3 h^2\right)+2 (n-1) h
\dot{\Omega}_m \right)+2 h \Omega_m \ddot{\Omega}_m\right)=0.
\end{align}
Now, transforming to redshift coordinates, one obtains
\begin{align}
h(z)^2&=\Omega_m(z)-\eta 
n^2 (z+1)^2 h(z)^2 \Omega_m(z)^{2 n-2} \Omega_m^{\prime2}\nonumber\\&-\epsilon n^3 (z+1)^3   h(z)^4\Omega_m(z)^{3 n-3} \Omega_m(z)^{\prime3}(z),
\end{align}
and
\begin{align}
\Omega_m(z)^4&-2 \eta  n^2 (z+1) h(z)
\Omega_m(z)^{2 n+1} \Big((z+1) \Omega_m(z) h'(z) \Omega_m^{\prime}(z)\nonumber\\&+h(z) \left((n-1) (z+1) \Omega_m^{\prime2}(z)+\Omega_m(z)
\left((z+1) \Omega_m^{\prime3}(z)-2 \Omega_m^{\prime}(z)\right)\right)\Big)\nonumber\\&- \epsilon n^3 (z+1)^2 h(z)^3 \Omega_m(z)^{3 n} \Omega_m^{\prime}(z) \Big(3
(z+1) \Omega_m(z) h'(z) \Omega_m^{\prime}(z)\nonumber\\&+h(z) \left(2 (n-1) (z+1)
\Omega_m^{\prime}(z)^2+\Omega_m(z) \left(2 (z+1)\Omega_m^{\prime2}(z)-\Omega_m^{\prime}(z)\right)\right)\Big)=0
\end{align}
In figure \eqref{fig4} we plot the Hubble and deceleration parameters for $n=2.5,3.5,4$ with $\eta=2.6,14.0,32.5$ and $\epsilon=-22.6,-269.4,-941.7$, for dotted, dot-dashed and dashed curves respectively. The red solid curve is the $\Lambda$CDM curve.
One can see from the figures that in the non-conservative case the early time behavior of the Hubble and deceleration parameters differ from the conservative case. The non-conservative case produce more acceleration in contrast to the conservative case. Also the Hubble parameter is greater than the $\Lambda$CDM curve which is also different from the conservative case. However the late time behavior is the same and can be fitted with observational data. One should note that in the non-conservative case, positive values of $n$ is also allowed and satisfy observational data. In figure \eqref{fig5} we have plotted the behavior of the density parameter $\Omega_m$ as a function of redshift.
 One can see from the figure that the density parameter coincides with observational data. However at early times, the matter density parameter becomes less than the standard $\Lambda$CDM model. This is in fact because of the non-conservative nature of the theory which makes the behavior different from $(1+z)^3$.
\section{Matter density perturbations}\label{matden}
In this section, we are going to consider dynamics of the matter density perturbations in the derivative matter coupling theory. For the sake of simplicity, we will only consider the dynamics of the scalar perturbation in the conservative case of the theory. After conformal transforming the time coordinate as $d\tau=dt/a$, the perturbed metric can be written as
\begin{align}
ds^2=a^2(\tau)\Big[-(1+2\Phi)d\tau^2+(1-2\Psi)d\vec{x}^2\Big],
\end{align}
where we used the Newtonian gauge $E=0=B$.
The perturbed energy momentum tensor can be defined as
\begin{align}
\delta T^0_0=-\delta\rho\equiv-\rho_0\delta,\quad \delta T^0_i=(1+c_s^2)\rho_0\partial_i v,\quad \delta T^i_j=\delta^i_jc_s^2\rho\delta.
\end{align}
Here, $\delta$ is the matter density perturbation defined as $\delta=\delta\rho/\rho_0$ and $v$ is the scalar mode of the velocity perturbation.
In the following, we will assume that the perturbed and unperturbed matter content of the Universe  have the equations of motion of the form $\delta p/\delta\rho=c_s^2=p_0/\rho_0$, where $c_s^2=0$.

With the above assumptions, one can obtain the energy momentum conservation equation to first order in perturbations as
\begin{align}
\delta^\prime-3\psi^\prime+\theta&=0,\label{matt1}\\
\theta^\prime+\mathcal{H}\theta-k^2\Phi&=0.\label{matt2}
\end{align}
where we have Fourier transfomed the perturbaed fields. Also, prime represents derivative with respect to $\tau$, $\mc{H}=a^\prime/a$, and $\theta=\nabla_i\nabla^i v$.
After combining equations \eqref{matt1} and \eqref{matt2}, one can obtain the dynamical equation for the matter density perturbation as
\begin{align}\label{denson}
\delta^{\prime\prime}+\mc{H}\delta^\prime-3\Psi^{\prime\prime}-\mc{H}\Psi^\prime+k^2\Phi=0.
\end{align}
In order to close the above equation one should obtain the scalar perturbations $\Psi$ and $\Phi$ from perturbed Einstein's equations. It can be easily seen that the $(ij), i\neq j$ components of the Einstein's equation \eqref{geom} give
$$\Psi=\Phi.$$
As a result, the $(00)$, $(0i)$ and $(ii)$ components of the Einstein's equation \eqref{geom} can be written as
\begin{align}\label{eq1}
3\mc{H}\Phi^\prime+3\mc{H}^2\Phi+k^2\Phi+\f{3\alpha}{2\kappa^2}\mc{H}\rho^2 f^{\prime2}\Big[\delta^\prime-3\mc{H}\left(1-\rho\f{f^{\prime\prime}}{f^\prime}\right)\delta+3\mc{H}\Phi\Big]=-\f{1}{4\kappa^2}a^2\rho\delta,
\end{align}
\begin{align}
k^2(\Phi^\prime+\mc{H}\Phi)-\f{3\alpha}{2\kappa^2}\rho^2f^{\prime2}\Big[k^2\mc{H}\delta+(\mc{H}^\prime-\mc{H}^2)\theta+3\mc{H}^2\theta\f{f^{\prime\prime}}{f^\prime}\Big]=-\f{1}{2\kappa^2}a^2\rho\theta,
\end{align}
\begin{align}
\Phi^{\prime\prime}&+3\mc{H}\Phi^\prime+2\Phi\mc{H}^\prime+\mc{H}^2
\Phi+\f{\alpha}{2\kappa^2}\rho^2f^{\prime2}\Big[\delta^{\prime\prime}+12\mc{H}\Phi^\prime-7\mc{H}\delta^\prime+3(\delta-\Phi)(5\mc{H}^2-2\mc{H}^\prime)+k^2\delta\Big]\nonumber\\
&+\f{3\alpha}{2\kappa^2}\rho^3f^{\prime2}\Big[3\rho\mc{H}^2\delta\f{f^{\prime\prime2}}{f^{\prime2}}+(2\delta\mc{H}^\prime+2\mc{H}\delta^\prime+6\mc{H}^2\Phi-14\mc{H}^2\delta)\f{f^{\prime\prime}}{f^\prime}+3\rho\mc{H}^2\delta\f{f^{\prime\prime\prime}}{f^\prime}\Big]=0,
\end{align}
where we have assumed $\beta=0$ for simplicity. In the sub-horizon limit $k\tau\gg1$, one can obtain $\Phi$ from \eqref{eq1} and the evolution equation of the matter density perturbation will be reduced to
\begin{align}
\delta^{\prime\prime}+\left(\mc{H}-\f{3\alpha}{2\kappa^2}\mc{H}\rho^2f^{\prime2}\right)\delta^\prime-\f32\mc{H}^2\left[1-\f{3\alpha}{2\kappa^2}\rho^2f^{\prime}\left(f^\prime-2\rho f^{\prime\prime}\right)\right]\delta=0.
\end{align}
Note that the above expression becomes equivalent to the standard $\Lambda$CDM for $\alpha=0$.
\begin{figure}[h]\label{fig6}
	\centering
	\includegraphics[scale=0.446]{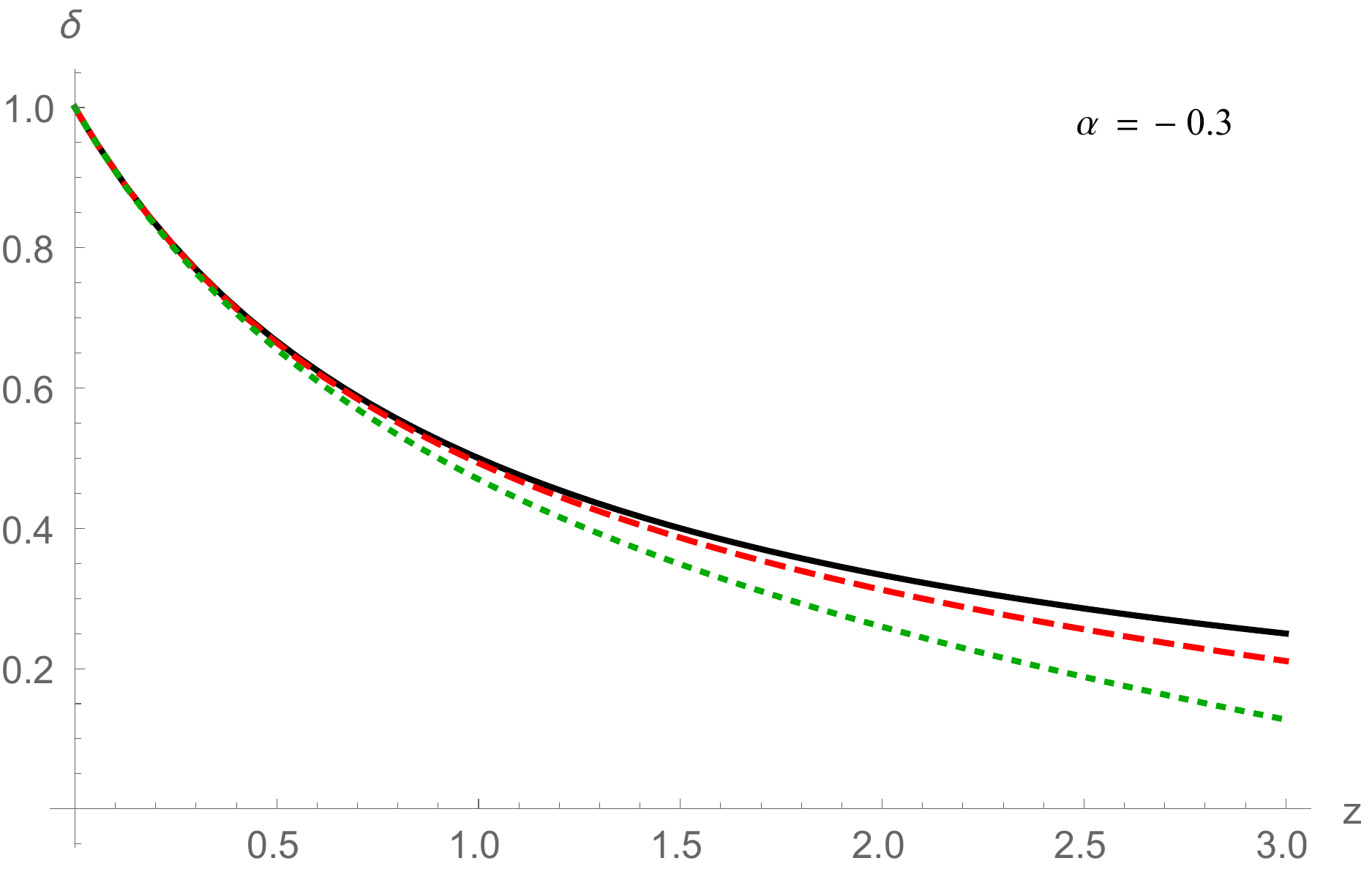}
	\includegraphics[scale=0.446]{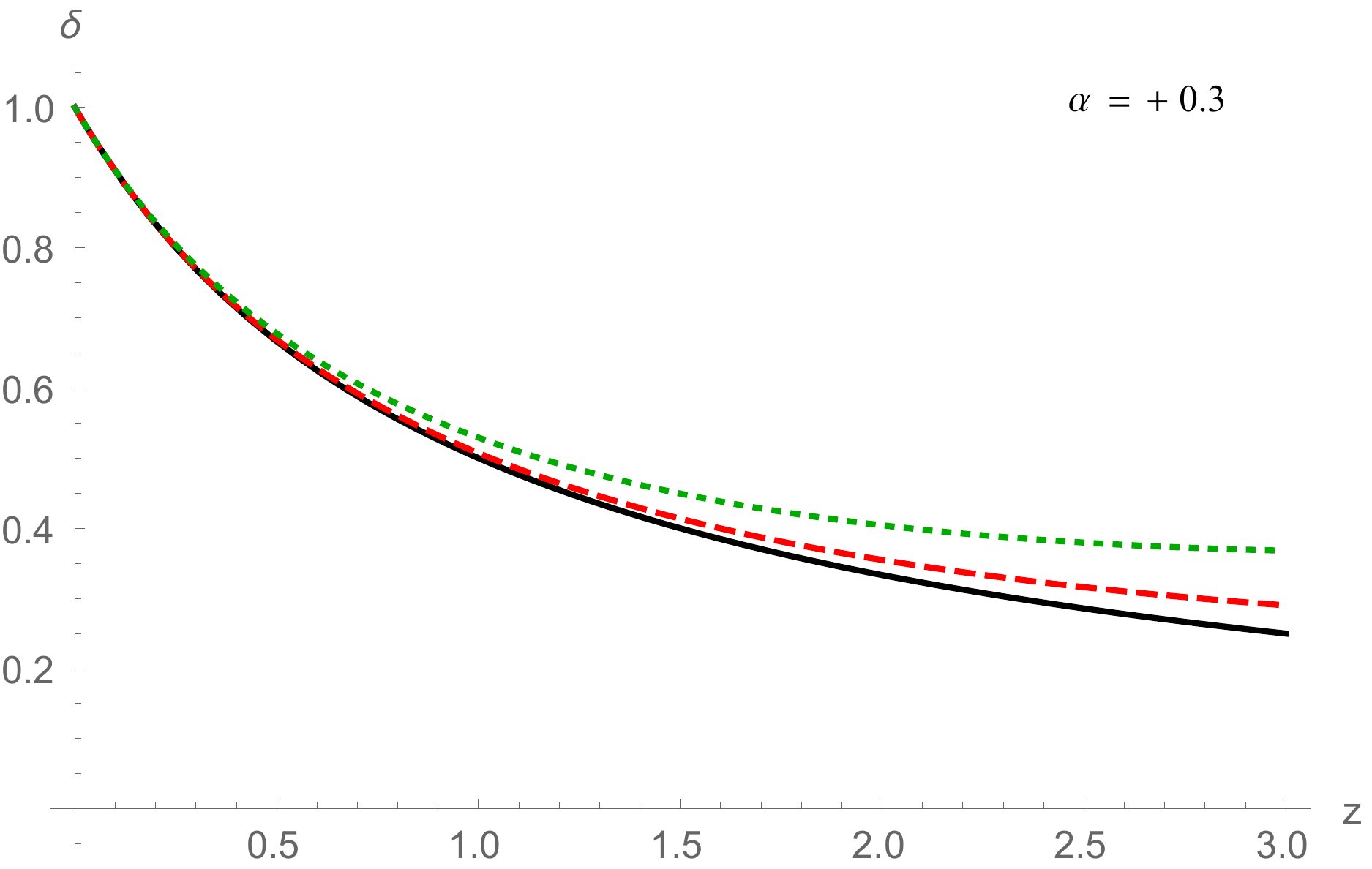}
	\caption{The variation of the matter density perturbation as a function of redshit $z$ for $n=0.2$ (dashed) and $n=-0.3$ (dotted) for $\alpha=0.3$ and $\alpha=-0.3$, for inverse exponential function $f\propto (L_m)^n$. The solid line indicates the $\Lambda$CDM curve.}
\end{figure}

In figure \eqref{fig5} we have plotted the evolution of the matter density perturbation as a function of the redshift $1+z=1/a$ for both signs of $\alpha$ in the case of $f=\gamma(-L_m/\rho_c)^n$. The solid curve represents the $\Lambda$CDM line corresponding to $\alpha=0$. One can see from the figures that the late time behavior of the model is in agreement with observational data. The early time behavior is different from the $\Lambda$CDM curve. In fact, positive values for $\alpha$ predicts larger density perturbations at early time, while negative values predict less density perturbations. This could be traced back to the fact that for positive values of $\alpha$ there would be more gravitational attractions from derivative matter couplings which shows itself in the early times. In contrast, negative values for $\alpha$ produce gravitational repulsion at early times.
\section{Conclusions and final remarks}\label{con}
In this paper, we have considered a theory containing derivative matter couplings. The matter interactions is expressed in terms of a new function of the matter Lagrangian $f=f(L_m)$. In order to write the action, remembering that $f$ is a scalar field, we chose a higher derivative scalar field interactions which is healthy in the sense of the Ostrogradski instability and also satisfy the gravitational wave observations \cite{gw}. The special class of the beyond Horndeski theory satisfy all the above requirements \cite{prls}. In this paper, we have considered the second and third Galileon interactions which is the simplest subclass of beyond Horndeski theory satisfying all the above constraints. Higher order interaction terms could also be added to the theory, but the qualitative behavior presented here does not change. Cosmological implications of the theory for three different functions $f$ corresponding to power-law, power-law-exponential and power-law-inverse-exponential is considered assuming that the Universe is filled by matter and radiation sources.
The energy momentum tensor is not conserved in general, signaling that the theory predicts an extra fifth force. The existence of the fifth force is a general behavior of the theories with matter/geometry coupling \cite{ketabeharko}. However, the present model has a property that on top of FRW space-time, the energy momentum tensor could be conserved. As a result the model, has a reach cosmology corresponding to both conservative and non-conservative cases. The conservative case made more acceleration and smaller Hubble parameter with respect to the $\Lambda$CDM at early times. The non-conservative case however, make the Universe to have less acceleration with respect to the $\Lambda$CDM at early times with greater Hubble parameter. Also, the non-conservative case predict less matter sources at early times compare to the $\Lambda$CDM case. The cosmological behavior of the model in both cases depends sharply on the values of the model parameters $\alpha$ and $\beta$ and also on the constant $n$ appearing in $f\propto L_m^n$. However, one can see that for the conservative case, the values of $n$ in the range $(-1,0)$ can satisfy late time observational data. 
In the case of non-conservative cosmology, all values of $n$ would be allowed.

In this paper, we have considered the evolution equation of the matter density perturbations in the dust dominated Universe. We have seen that in the deep sub-horizon limit with $k\tau\gg1$, the behavior of density parameter coincides at late time to the standard $\Lambda$CD model. However, at early times, the behavior depends on the value of $\alpha$. Positive values predict more density and negative values predict less density than the standard $\Lambda$CDM model. However, the qualitative behavior of the matter density perturbation does not depend on the value of $n$ in $f\propto L_m^n$. More detailed analysis of the matter density perturbation would be the scope of an independent work.

\end{document}